\documentclass[12pt]{article}

\pdfoutput=1

\usepackage[top=80pt,bottom=85pt,left=85pt,right=85pt]{geometry}
\usepackage{amsmath,amssymb,graphicx,float,color,tikz,cite,savesym,wasysym}
\usepackage{caption}
\usepackage{subcaption}
\usepackage[debug,pageanchor=false]{hyperref}
\definecolor{link}{rgb}{.8,.15,.1}
\hypersetup{colorlinks=true,linkcolor=link,citecolor=link,urlcolor=link,linktocpage}
   
\newcommand{\R}{\mathbb{R}}

\renewcommand{\Re}{\mathrm{Re}}    
\renewcommand{\Im}{\mathrm{Im}}  

\newcommand{\dd}{\mathrm{d}}
\newcommand{\e}{\mathrm{e}}
\newcommand{\ii}{\mathrm{i}}
\newcommand{\vol}{\mathrm{vol}}

\makeatletter
\@addtoreset{equation}{section}
\makeatother

\begin{document}

	\begin{titlepage}

	\begin{center}

	\vskip .5in 
	\noindent

	{\Large \bf{Breaking supersymmetry with pure spinors}}

	\bigskip\medskip

	 Andrea Legramandi and  Alessandro Tomasiello \\

	\bigskip\medskip
	{\small 
 Dipartimento di Fisica, Universit\`a di Milano--Bicocca, \\ Piazza della Scienza 3, I-20126 Milano, Italy \\ and \\ INFN, sezione di Milano--Bicocca
	
		}

	\vskip .5cm 
	{\small \tt andrea.legramandi, alessandro.tomasiello@unimib.it}
	\vskip .9cm 
	     	{\bf Abstract }
	\vskip .1in
	\end{center}

	\noindent

For several classes of BPS vacua, we find a procedure to modify the PDEs that imply preserved supersymmetry and the equations of motion so that they still imply the latter but not the former. In each case we trace back this supersymmetry-breaking deformation to a distinct modification of the pure spinor equations that provide a geometrical interpretation of supersymmetry. We give some concrete examples: first we generalize the Imamura class of Mink$_6$ solutions by removing a symmetry requirement, and then derive some local and global solutions both before and after breaking supersymmetry.

	\noindent

	\vfill
	\eject

	\end{titlepage}

\tableofcontents

\section{Introduction} 
\label{sec:intro}

If supersymmetry exists, it appears it is broken at high energy scales. In string theory it does play an important role, but nothing prevents it from being spontaneously broken at the Planck scale. 

Finding supersymmetric solutions, however, is still a lot easier than finding non-supersymmetric ones. This is in part because the BPS equations provide a first-order system of partial differential equations (PDEs). Moreover, these equations often have compelling geometrical interpretations. These are revealed for example in the $G$-structure formalism (starting with \cite{strominger,gauntlett-pakis}), and in more complicated cases by generalized complex geometry methods \cite{gmpt2}. 

Indeed the latter provide a system of ``pure spinor equations'' \cite{gmpt2} that partially reduce finding the most general Minkowski or AdS supersymmetric vacuum solution to a geometrical problem. For Minkowski, for example, a condition that emerges is that the internal space be ``generalized complex'', an umbrella concept that contains complex and symplectic manifolds. This is not enough to find a solution, but provides a convenient first step. 

It would be very useful to have similar methods for non-supersymmetric solutions. Roughly speaking, the hope would be to modify the aforementioned first-order geometrical systems, to obtain a new one that still implies the equations of motion (EoMs), when supplemented with the Bianchi identities, but which is no longer equivalent to the BPS conditions. This hope is partially inspired for example by the idea of fake superpotential in lower-dimensional theories. 

This has been attempted in the past; for example, \cite{lust-marchesano-martucci-tsimpis} has parameterized the most general deformation of the BPS system of \cite{gmpt2}, and has imposed the condition that such a deformation implies the EoMs. This approach appears promising; unfortunately the resulting constraints on the deformations are rather intricate, to the point where it is currently a bit unpractical. 

In this paper we reexamine the problem by looking at some specific classes of increasing complexity. By a ``class'' we mean a set of supersymmetric solutions, where the metric, dilaton and fluxes have been fixed, up to solving a system of PDEs. For each class we provide a way to modify the PDEs by a supersymmetry-breaking parameter, so that the new system of PDEs still implies the EoMs but not the BPS equations. 

Let us give an example. The Imamura class \cite{imamura} is a set of supersymmetric Mink$_6\times M_4$ solutions, with fluxes $F_0$, $F_2$ and $H$. All fields are parameterized by a single function $S$, obeying a single PDE\footnote{This is generically true, namely for $F_0\neq 0$; for $F_0=0$, there are two functions, obeying two PDEs.} 
\begin{equation}\label{eq:ima-intro}
	\triangle_3 S + \frac12 \partial^2_z S^2 =0 \,,
\end{equation}
valid away from sources.\footnote{The PDE is second-order because it is generated by acting with the Bianchi identity on the BPS equations.} Here $\triangle_3$ is a Laplacian on three coordinates $x_i$, that together with $z$ span the internal space $M_4$. Originally in \cite{imamura} this PDE was derived assuming SO(3) invariance in the $x_i$ (or in other words dependence on $r=\sqrt{x_i x_i}$ only). Already with this assumption this class is rather interesting: for example it was found in \cite{bobev-dibitetto-gautason-trujien,macpherson-t} that in a limit it generates the AdS$_7$ solutions of IIA \cite{afrt}. 

In this paper we check that (\ref{eq:ima-intro}) implies the BPS equations even without this symmetry assumption. When liberated from this artificial constraint, this class reveals itself to be even richer than previously thought; applying some standard techniques we find a lot of local solutions, and at least one compact solution (building on \cite{janssen-meessen-ortin,imamura,blaback-janssen-vanriet-vercnocke,blaback-vanderwoerd-vanriet-williams}) where $M_4$ has the topology of $T^4$, and contains O8-planes, and D6-branes, all localized and back-reacting on the geometry. 

Coming now to our supersymmetry-breaking technique, on the Imamura class we obtain a new PDE that reads
\begin{equation}\label{eq:lima-intro}
	\triangle_3 S + \frac12 \partial^2_z S^2+ c(c - 2 \partial_z S) =0\,.
\end{equation}
For $c=0$ this reduces to (\ref{eq:ima-intro}). For $c\neq 0$, we get a deformation of the Imamura equation which still implies all the EoMs. While not all the techniques that work with (\ref{eq:ima-intro}) still succeed with (\ref{eq:lima-intro}), at least one does, and again gives a class of compact solutions, this time non-supersymmetric.

The Imamura class is a rare case where all equations can be reduced to a single one. (Another such class is ${\mathcal N}=2$ AdS$_5$ solutions in eleven dimensions \cite{lin-lunin-maldacena}, where everything is reduced to a single Toda equation, which in fact has some similarities to (\ref{eq:ima-intro}).) With other classes, things are not so simple and one in general gets a system which is usually let's manageable then (\ref{eq:ima-intro}). Still, we manage to obtain similar results as (\ref{eq:lima-intro}) even in more complicated cases. While we do not reach the point where we can give an algorithmic procedure in full generality, we find patterns that we think might be useful for further investigation. 

In particular, in a certain sense we will specify better below, we are able to reverse-engineer our supersymmetry-breaking results such as (\ref{eq:lima-intro}) to a specific modification of the pure spinor equations of \cite{gmpt2}. Namely, if we keep fixed the ansatz for the pure spinors in a given supersymmetric class, the modification of the RR fluxes (and hence the Bianchi identities) can be inferred from the usual pure spinor equation. 

Although promising, our method has currently some limitations. First of all, as we mentioned, while we do see some emerging patterns, at the present stage our procedure is not fully algorithmic, and requires some guesswork which we have carried out case by case.  Second, in this paper we only looked at Minkowski solutions. In a next stage it would be natural to try to apply it to solutions with cosmological constant; for example for AdS solutions it would be interesting to see whether some of the supersymmetric vacua one can obtain by consistent truncation (for example the non-BPS AdS$_7$ solutions noticed in \cite{passias-rota-t})  can be generated in a similar way. Even better would be to be able to change the cosmological constant in the process, perhaps generating dS vacua by breaking supersymmetry in a Minkowski or AdS solution. Finally, the procedure suffers from a familiar problem: if we try to embed these supergravity vacua in string theory, any surviving moduli in the non-BPS solutions one generates this way are likely to get a non-zero potential from quantum corrections. In the example (\ref{eq:lima-intro}), there is a new parameter $c$ that naively would even provide a new modulus; however in general this appears in a flux component, and one can expect it to be discretized by flux quantization. Let us stress that our method does not address this modulus problem, and it should be regarded as a solution-generating technique in supergravity. 

Besides the application to finding solutions, these methods can be conceptually useful from several points of view. The pure spinor equations for BPS solutions have an interpretation in terms of calibrations; one can thus expect that non-supersymmetric vacua obtained with pure spinor methods might still have calibrated branes. This might also imply good stability properties, since branes often provide non-perturbative decay channels for non-supersymmetric vacua.

In section \ref{sec:susy} we review the pure spinor equations; then we use them to rederive and generalize some supersymmetric classes of solutions, including the Imamura class we mentioned around (\ref{eq:ima-intro}). In section \ref{sec:susy-br} we will explain our strategy for breaking supersymmetry, applying it to the previously introduced BPS classes. Finally in section \ref{sec:ex} we will see some examples, first in the BPS Imamura class and then in its supersymmetry-breaking counterpart.


\section{Supersymmetry} 
\label{sec:susy}

We start in this section with the usual preliminaries, in particular reviewing very quickly the pure spinor method in section \ref{sub:psp}. In subsequent subsections we will get more specific and present some interesting subclasses of solutions, whose supersymmetry we will break later in the paper.

As we anticipated in the introduction, we are interested in Mink$_d\times M_{10-d}$ solutions with $d \ge 4$ in type II supergravity. The case $d>4$ can be seen as a particular case of the case $d=4$, for which the metric reads 
\begin{equation}
\label{eq:metric_mink4}
\dd s^2_{10} = e^{2A}\dd s^2(\R^{1,3}) + \dd s^2(M_6)\,.
\end{equation} 
Solutions with higher-dimensional external space are included by further splitting $M_6$:
\begin{equation}
\label{eq:higher_mink}
\dd s^2(M_6) = e^{2A}\dd s^2(\R^{d}) + \dd s^2(M_{6-d}) \, .
\end{equation}

In order to preserve the Poincar\'e isometries of Mink$_4$, we have to assume that the warping function $A$ only depends on the six-dimensional manifold $M_6$ (or on $M_{6-d}$, if we are considering the case \eqref{eq:higher_mink}). Moreover, we have to require that fluxes have no legs along $\R^{1,3}$ except for the volume form:
\begin{equation}
F= f + e^{4A} \vol(\R^{1,3}) \wedge *_6 \lambda f\,.
\end{equation}
$f$ is a form on $M_6$ only, and $\lambda$ is a sign defined by $\lambda=(-)^{k(k-1)/2}$, where $k$ is the form degree. Notice that we are adopting the democratic formalism \cite{democratic}, in conventions however where $F= * \lambda F$. We also have to impose that the NSNS three form $H$ is strictly a form on $M_6$. The Bianchi identities away from sources read $\dd_H F=0$, $\dd H=0$,
where $\dd_H\equiv \dd - H \wedge$.

For supersymmetry, we also need to consider the fermionic parameters $\epsilon^{1,2}$. Poincar\'e symmetry requires that they factorize in terms of a constant spinor $\zeta$ on $\R^{1,3}$ and a couple of spinors $\eta^{1,2}$ on $M_6$
\begin{equation}\label{eq:eps-zeta}
\epsilon^{1,2} = \zeta_+ \otimes \eta^{1,2}_\pm + \zeta_- \otimes \eta^{1,2}_\mp \, ,
\end{equation} 
where the chirality $+/-$ depends if we are in type IIA/IIB theory respectively, and we chose the charge conjugation so that $\zeta_+ = \overline{(\zeta_-)}$ and $\eta^2_+= \overline{(\eta^2_-)}$.

However, in this paper we will not use directly the spinorial formalism, but we will appeal to the pure spinor method, which allows to reformulate the problem of finding four-dimensional vacuum solutions. 

\subsection{Pure spinors} 
\label{sub:psp}

The pure spinor method \cite{gmpt2,gmpt3} provides a way to express the BPS conditions in terms of forms; here we will give a lightning review of it, to set the stage for our supersymmetry-breaking modification in the next sections. 

In general a spinor is called pure if it annihilator in the Clifford algebra is the largest possible, i.e.~if it has half the dimension of spacetime. Actually the pure spinors we need are polyforms, namely formal sums of various differential forms of different degrees. Indeed polyforms can be regarded as spinors for the ``doubled'' Clifford algebra, in our case $\mathrm{Cl}(6,6)$. Starting from the internal spinors $\eta^{1,2}$ appearing in (\ref{eq:eps-zeta}), we can form the bispinors
\begin{equation}
\label{eq:pure_spinors_and_spinors}
\Phi_+ = \eta^1_+ \otimes \eta^2_+ \, , \qquad   \Phi_+ = \eta^1_+ \otimes \eta^2_- \, .
\end{equation}
Using Fierz identity and the Clifford map $\gamma^{m_1 \dots m_k} \mapsto dx^{m_1}\wedge \ldots \wedge dx^{m_k}$, these can be interpreted as polyforms, and can be shown to be pure, basically because the $\eta^{1,2}_+$ are already pure as $\mathrm{Cl}(6)$ spinors. The $+$ ($-$) label on $\Phi_\pm$ indicates that all the forms appearing in them have even (odd) degree. 

The BPS equations are originally written in terms of the spinors $\eta^{1,2}$, but it is possible to reformulate them completely in terms of the $\Phi_\pm$. While not all  differential forms are tensor products of two spinors as in (\ref{eq:pure_spinors_and_spinors}), that requirement is equivalent to the compatibility conditions
\begin{equation}\label{eq:comp}
\begin{split}
&(\Phi_-, \gamma^m \cdot \Phi_+) = (\Phi_-, \Phi_+ \cdot \gamma^m) = 0 \qquad \forall m\,, \\
&(\overline{\Phi}_-, \gamma^m \cdot \Phi_+) = (\overline{\Phi}_-, \Phi_+ \cdot \gamma^m) = 0 \qquad \forall m\,;\\
& (\Phi_+ , \overline{\Phi}_+) = (\Phi_- , \overline{\Phi}_-) = - \frac{\ii}{8} \vol(M_6) \, .
\end{split}
\end{equation}
Here $(\alpha , \beta)\equiv (\alpha \wedge \lambda(\beta))_6$ is the six-dimensional Chevalley--Mukai pairing and $\cdot$ is the Clifford product acting on a differential form: $\gamma^m \cdot = \dd x^m \wedge + \iota^m$, $\cdot \gamma^m=\pm (\dd x^m \wedge - \iota^m)$. 
In other words, if these constraints are satisfied, there exist $\eta^{1,2}_\pm$ such that $\Phi_\pm$ can be written as (\ref{eq:pure_spinors_and_spinors}). 

The purity requirement and the compatibility constraints (\ref{eq:comp}) are algebraic conditions, whose general solution is known:\footnote{Actually it is possible to consider also the twisted version of these pure spinors, but since a $B$-transformation sends compatible pairs in compatible pairs, we can set $B=0$.}
\begin{equation}
\label{eq:dynamicSU(2)}
\Phi_+ = \frac{1}{8} \e^{\frac{1}{2} E_3 \wedge \overline{E}_3} \wedge (k_\parallel\e^{- \ii j} + \ii k_\perp \omega) \, , \qquad \Phi_- = \frac{1}{8}  E_3 \wedge (\overline{k}_\perp \e^{- \ii j} - \ii \overline{k}_\parallel  \omega) \, .
\end{equation}
where $\{E_1,E_2,E_3\}$ is a local complex vielbein and $|k_\perp|^2 + |k_\parallel|^2=1$, and $j,\omega$ define the SU(2)-structure
\begin{equation}
\label{eq:SU(2)-vielbein}
j = \frac{\ii}{2} (E_1 \wedge \overline{E}_1+E_2 \wedge \overline{E}_2) \, , \qquad \omega =  E_1 \wedge E_2 \, .
\end{equation}
Two interesting cases, which will play a role in this paper, are the SU(3)-structure case $k_\perp = 0$, $k_\parallel= - \ii $, and the static SU(2)-structure case $k_\parallel= 0$, $k_\perp = 1$. (In terms of the $\eta^{1,2}$, these correspond to the case where they are proportional and orthogonal, respectively.)

Given this reformulation of the pair of spinors $\eta^{1,2}$ in terms of the pair of pure polyforms $\Phi_\pm$, one can rewrite the BPS conditions in terms of the $\Phi_\pm$ \cite{gmpt2}:
\begin{subequations}
\label{pure_spinors}
\begin{align}
&\dd_H ( e^{3A-\phi} \Phi_\pm) = 0 \, ,\label{pure_spinors1}\\
&\dd_H( e^{2A-\phi} \Re\Phi_\mp) = 0 \, ,\label{pure_spinors2} \\
&\dd_H ( e^{4A-\phi}\Im\Phi_\mp) = \frac{\e^{4A}}{8} *_6 \lambda(f) \, , \label{pure_spinors3}
\end{align}
\end{subequations} 
where the upper sign is for type IIA while the lower one for type IIB. (For simplicity, we have restricted ourselves to the case where the spinor norms are equal, $|\eta^1_+|=|\eta^2_+|$; this is believed to be necessary for compact solutions, for example.) The system of pure spinor equations (\ref{pure_spinors}) is equivalent to the BPS equations; moreover, together with the Bianchi identities for the RR fluxes, it implies all the other equations of motion. 

For this system it was given an interpretation in terms of $d=4$, ${\mathcal N}=1$ supergravity in \cite{koerber-martucci-ten-four}. By introducing a suitable superpotential $W$, it was found there that (\ref{pure_spinors1}) is an F-term equation resulting from varying $W$ with respect to the moduli ${\mathcal T}$ corresponding to deformations of $\Phi_\mp$, (\ref{pure_spinors3}) results from varying $W$ with respect to the moduli ${\mathcal Z}$ of $\Phi_\pm$, and finally (\ref{pure_spinors2})  is a D-term equation. Summarizing:
\begin{equation}\label{eq:DW}
	(\ref{pure_spinors1})= \partial_{\delta \Phi_\mp} W \, ,\qquad (\ref{pure_spinors2}) = D \, ,\qquad (\ref{pure_spinors3}) = \partial_{\delta \Phi_\pm} W \, .
\end{equation}


\subsection{The Imamura class} 
\label{sub:imamura}

Let us now apply the formalism of the previous section to a particular case of $\frac{1}{4}$-BPS system of intersecting branes in type IIA supergravity. Specifically, we will consider a generalization of the localized D6-D8-NS5 setup described in \cite{imamura}\footnote{The system in \cite{imamura} is itself a generalization of \cite[Sec.~3]{janssen-meessen-ortin}.} in which we drop the spherical symmetry ansatz on the internal space. (This generalization was also considered in \cite{blaback-janssen-vanriet-vercnocke}.)

The class actually describes an $\R^{1,5} \times M_4$ space-time. The pure spinor equations (\ref{pure_spinors}) assume a four-dimensional Minkowski space; if we want two extra flat directions, we have to implement \eqref{eq:higher_mink} for $d=2$. This can be done in various ways, for example by imposing that one of the forms of the complex vielbein is locally defined as 
$\e^A (\dd y_1 + \ii \, \dd y_2)$.
Alternatively, we could use the system in \cite{lust-patalong-tsimpis}, which is directly the analogue of (\ref{pure_spinors}) for Mink$_6$ solutions.

Let us now call $E_2 = w = w_1 + \ii w_2$ and $E_3 = v = v_1 + \ii v_2$, where $\{w_1,w_2,v_1,v_2\}$ is a complex vielbein on $M_4$. Inserting in \eqref{eq:dynamicSU(2)}  $k_\parallel= 0$, $k_\perp = 1$ and 
\begin{equation}\label{eq:E1y}
	E_1=\e^A (\dd y_1 + \ii \, \dd y_2)
\end{equation}
inside \eqref{pure_spinors} we get the two-form conditions
\begin{equation}
\dd (\e^{4A - \phi} w) = \dd (\e^{4A - \phi} v_2)= \dd (\e^{2A - \phi} v_1) = 0\,,
\end{equation}
which can be solved introducing local coordinates:
\begin{equation}
w = \e^{-4A + \phi} (\dd x_1+ \ii \, \dd x_2) \, , \qquad v = \e^{-2A + \phi} \dd z+ \ii \, \e^{-4A + \phi} \dd x_3 \, .
\end{equation}
This means that the metric can be written as
\begin{equation}
\label{eq:ima_metric}
\dd s^2_{10} = \e^{2A} \dd s^2 (\R^{1,5})+ \e^{-4A+2\phi} \dd z^2+ \e^{-8A+2\phi} \dd s^2(\R^3) \, ,
\end{equation}
where $\R^3$ is spanned by $\{x_1,x_2,x_3\}$. 

If we introduce $S\equiv e^{-4A}$ and $K\equiv \e^{-6A+2\phi}$, we now recover the metric in \cite[(2.2)]{imamura}. The metric factor become $S^{-1/2}$, $K S^{-1/2}$, $K S^{1/2}$; the dilaton reads $\e^{\phi}= K^{1/2} S^{-3/4}$. We recognize the structure one would expect for an NS5-D6 brane system brane configuration; moreover, since as in \cite{imamura} we will include a non-zero $F_0$, one might expect the possibility of a D8 transverse to $z$. We summarize this in table \ref{table:D6-D8-NS5}. 
\begin{table}[h]
	\centering
	\begin{tabular}{ccccccccccc} 
		\hline
		 & $0$ & $1$ & $2$ & $3$ & $y_1$ & $y_2$ & $z$ & $x_1$ & $x_2$ & $x_3$      \\ 
		\hline
		 NS5 & $\circ$ &$\circ$ &$\circ$ & $\circ$ & $\circ$ & $\circ$ &  &  &  &    \\ 
		 D6 & $\circ$ &$\circ$ &$\circ$ &$\circ$ & $\circ$ & $\circ$ & $\circ$  &  &  &    \\ 
		 D8 & $\circ$ &$\circ$ &$\circ$ &$\circ$ & $\circ$ & $\circ$ &  & $\circ$  & $\circ$  & $\circ$    
		\\ \hline
	\end{tabular}
	\caption{Localized D6-D8-NS5 brane system.}
	\label{table:D6-D8-NS5}
\end{table}

Turning now to the higher degree equations in (\ref{pure_spinors}), we see that they define the fluxes in terms of $A$ and $\phi$ as follows:
\begin{subequations}
\label{Imamura_fluxes}
\begin{align}
&H= - \frac{1}{2} \epsilon^{ijk} \dd z \wedge \dd x_i \wedge \dd x_j \partial_k \e^{-6A+2\phi} + \vol(\R^3)  \partial_z \e^{-10 A+2\phi} \, , \label{Imamura_fluxesH} \\
&F_0 =2 \e^{-2 \phi } \partial_{z} \e^{2 A} \label{eq:Imamura-F0} \, ,\\
&F_2 =  \frac{1}{2} \epsilon^{ijk} \dd x_i \wedge \dd x_j \partial_{x_k} \e^{-4A} \, , \label{eq:Imamura-F2}\\
&F_4=0 \, . \label{eq:Imamura-F4}
\end{align}
\end{subequations}
We took advantage of the explicit form of the metric to explicitly compute the Hodge dual in \eqref{pure_spinors3}.

Notice that the functions $A$, $\phi$ can depend on all four coordinates; in this respect our class is more general than the original one in \cite{imamura}, where both functions were taken to depend on $z$ and on a radial coordinate $r\equiv (x_i x_i)^{1/2}$, and so there was an additional $\mathrm{SO}(3)$ symmetry we are not assuming here.

The original Imamura class with this $\mathrm{SO}(3)$ also emerged in \cite[section 4.1]{macpherson-t}, starting from a SU(2)-structure bi-spinor ansatz which was reduced to an identity structure. This reduction occurs when we impose that one of the complex directions which define $j,\omega$ in \eqref{eq:SU(2)-vielbein} is actually part of the external space $\R^{1,5}$, as in (\ref{eq:E1y}). This is not the only way to get an identity structure assuming a six-dimensional Minkowski space; however it was found in \cite{l-macpherson} that these cases give rise to parametric deformations of the usual Imamura solution which can be generated by chains of dualities. Since none of these dualities requires the presence of the $\mathrm{SO}(3)$ symmetry in the co-dimensions of the D6-brane, we don't expect the discussion is much different dropping this assumption.

So far we have only imposed the pure spinor equations; we now turn to the Bianchi identities. Since the Romans mass must be a constant 
\begin{equation}
	F_0=m\,,
\end{equation}
we have two different cases, depending on whether $m$ is zero or non-zero.

\subsubsection{Massive case}

In this case we can use (\ref{eq:Imamura-F0}) to write $\phi$ as a function of $A$:
\begin{equation}
\label{eq:phi_susy}
\e^{2\phi} = \frac{2}{m}\partial_{z}\e^{2 A}.
\end{equation}
This redefinition combined with the Bianchi identity for $F_2$ gives the Imamura PDE
\begin{equation}
\label{eq:Ima}
\Delta_3 \e^{-4A}+\frac{1}{2}\partial_{z}^2 \e^{-8A}=0 \, ,
\end{equation} 
where $\Delta_3= \partial_{x_1}^2+\partial_{x_2}^2+\partial_{x_3}^2$. The Bianchi identity for $H$ gives $\partial_z$ of (\ref{eq:Ima}), and hence is automatically implied by it.

\subsubsection{Massless case}

If $F_0=0$, (\ref{eq:Imamura-F0}) implies $\partial_{z}e^{2 A} = 0$. The Bianchi identity for $F_2$ now imposes
\begin{equation}
\Delta_3 \e^{-4A} = 0 \, .
\end{equation}
In this case the Bianchi identity for $H$ is not automatically implied by the one for $F_2$, so it should be checked independently. One finds that it imposes: 
\begin{equation}
\Delta_3 \e^{-6A + 2 \phi}+\e^{-4A}\partial_{z}^2 \e^{-6A + 2 \phi} = 0 \, ,
\end{equation}
which is a Youm-like condition \cite{youm}.


\subsection{A larger IIA system} 
\label{sub:largerIIA}

In \cite[App.~C]{l-macpherson} it was proven that the Imamura solution can be derived by a more general $\R^{1,3} \times S^2$ solution of type IIA supergravity by imposing two isometric directions and T-dualizing along them. In this section, we will generalize that result relaxing the rotational symmetry ansatz, as done in the previous section. This will provide a generalization of the class in \cite[App.~C]{l-macpherson} that extends the class of subsection \ref{sub:imamura}.

Since without any ansatz on the internal space it is impossible to get an identity structure, we start by defining directly the six-dimensional complex vielbein in terms of local coordinates as follows:
\begin{equation}
E_1 = -\e^{-A} (\dd y_1 + \ii \, \dd y_2 ) \, , \quad E_2 = -\e^{-2A+ \phi} (\dd x_2 - \ii \, \dd x_3 ) \, , \quad  E_3 = \e^{-2A+ \phi} (\dd x_1 + \ii \, \e^{2A+ \mu} \dd z ) \, .
\end{equation}
The functions are now chosen so that some of the pure spinor equations are automatically satisfied.
This gives the ten-dimensional metric
\begin{equation}
\label{eq:metIIA_noS2}
\dd s^2 = \e^{2A}\dd s^2 (\R^{1,3}) + \e^{-4A + 2 \phi}  \dd s^2(\R^3)+ \e^{-2A} \dd s^2(\R^2) + \e^{2 \phi + 2 \mu} \dd z^2 \, ,
\end{equation}
where $y_i \in \R^2$ and $x_i \in \R^3$. Notice that if we impose that $\partial_{y_i}$ are two Killing directions and we T-dualize along them we exactly get the metric \eqref{eq:ima_metric}, thus recovering the Imamura class.

The easiest generalization of \cite[App.~C]{l-macpherson} is obtained by imposing that $H$ has legs only along $\R^3$ and $z$. It is now particularly easy to solve the pure spinor equations \eqref{pure_spinors} with the ansatz $k_\parallel= 0$, $k_\perp = 1$. 
We define
\begin{equation}
\label{eq:phi_f}
f = \e^{-2A+2\phi-\mu}\,;
\end{equation}
supersymmetry then imposes that $f$ just depends on $(z,x_1,x_2,x_3)$, while $\mu = \mu(z,y_1,y_2)$.
The fluxes read:
\begin{subequations}
	\label{susyIIA_fluxes}
	\begin{align}
	&H= - \frac{1}{2} \epsilon^{ijk} \dd z \wedge \dd x_i \wedge \dd x_j \partial_k f + \vol(\R^3) \e^{-\mu}  \partial_z (\e^{-4A-\mu} f) \, , \label{eq:susyIIA_H} \\
	&F_0 =0 \, ,\\
	&F_2 = \left( \partial_{y_2}\e^{\mu} \dd y_1 -\partial_{y_1}\e^{\mu} \dd y_2 \right) \wedge \dd z - f^{-1} \partial_z \e^{-4A} \vol(\R^2)  \, ,\\
	&F_4= f \vol(\R^3)\wedge  \left( \partial_{y_2}\e^{-4A-\mu} \dd y_1 - \partial_{y_1}\e^{-4A-\mu} \dd y_2 \right) + \frac{1}{2} \epsilon^{ijk} \dd x_i \wedge \dd x_j \wedge \vol(\R^2) \partial_{x_k} \e^{-4A} \, .
\end{align}
\end{subequations}
The Bianchi identities reduce to 
\begin{subequations}
	\label{eq:largerIIA_BI1}
	\begin{align}
	&\partial_{x_i} \left( f^{-1} \partial_z \e^{-4A} \right) =0 \, , \\
	&\partial_{y_i} \left(\e^{-\mu} \partial_{z} (f \e^{-4A-\mu}) \right) =0 \, ,
	\end{align}
\end{subequations}
which restrict the functional dependence of the various function in play, plus the PDEs
\begin{subequations}
	\label{eq:largerIIA_BI2}
	\begin{align}
	&\triangle_2 \e^{\mu} +\partial_{z} \left( f^{-1} \partial_z \e^{-4A} \right) =0 \, , \\
	&\triangle_3 f + \partial_z \left(\e^{-\mu} \partial_{z} (f \e^{-4A-\mu}) \right) = 0 \, , \\
	&f \triangle_2 \e^{-4A-\mu} + \triangle_3 \e^{-4A} + f^{-1} \partial_{z}\e^{-4A}  \left(\e^{-\mu} \partial_{z} (f \e^{-4A-\mu}) \right) = 0 \, ,
	\end{align}
\end{subequations}
where $\triangle_2$ and $\triangle_3$ are the Laplacian on $\R^2_y$ and $\R^3_x$ respectively.



\section{Supersymmetry breaking} 
\label{sec:susy-br}

In this section we will see that in some circumstances it is possible to extend the pure spinor equations \eqref{pure_spinors} to non-supersymmetric cases. 

\subsection{Breaking supersymmetry in the Imamura class} 
\label{sub:im-sb}

Let us start by considering the Imamura class we derived in section \ref{sub:imamura}. 

Since we would like to extend this class to a non-supersymmetric setting, let us relax some of the conditions we found by imposing the pure spinor equations. In principle it would be possible to relax all the equations and move to a completely different solution; however, to fix ideas, we will keep the definition of the RR-fluxes and the metric, and not impose anything else. We also require that $F_0 = m \neq 0$. So far the background is simply given by the usual Imamura metric (\ref{eq:ima_metric}),
\begin{equation}
\dd s^2_{10} = \e^{2A} \dd s^2 (\R^{1,5})+ \e^{-4A+2\Phi} \dd z^2+ \e^{-8A+2\Phi} \dd s^2(\R^3) \, ,
\end{equation}
with RR fluxes as in (\ref{eq:Imamura-F0})--(\ref{eq:Imamura-F4}):  
\begin{equation}
\label{eq:Ima-flux_repeat}
	\begin{split}
	&F_0 =m \, , \qquad F_4 = 0 \, ,\\
	&F_2 =  \frac{1}{2} \epsilon^{ijk} \dd x_i \wedge \dd x_j \partial_{x_k} \e^{-4A} \, . \\
	\end{split}
\end{equation}

But instead of defining $H$ as in \eqref{Imamura_fluxesH}, we take the Bianchi identity for $F_2$ as its definition:
\begin{equation}
\label{eq_H_def_non_susy_Ima}
H = \frac{1}{m}\dd F_2 \, .
\end{equation}
The expression for $H$ can be simplified thanks to the external Einstein equation, which reads
\begin{equation}
\label{eq:Eins_0_0}
\Delta_3 \e^{-4A} +  \e^{-4A} \partial_z^2 \e^{-4A} + \e^{-12A+2 \phi}m^2 = 0 \, .  
\end{equation}
Using this equation in \eqref{eq_H_def_non_susy_Ima} one gets
\begin{equation}
H= \frac{1}{2m} \epsilon^{ijk} \dd z \wedge \dd x_i \wedge \dd x_j \partial_{x_k} \partial_z \e^{-4A} - \frac{1}{m} \vol(\R^3) ( \e^{-4A} \partial_z^2 \e^{-4A} + \e^{-12A+2 \phi}m^2) \, . 
\end{equation}
This definition of $H$ turns out to be particularly useful when one has to compute the equation of motion for the B-field:
\begin{equation}
\label{eq:B_EoM_non_susy_Ima}
\dd (\e^{-2 \phi} * H) = F_0 F_8 \, .
\end{equation}
Thanks to the fact that we can explicitly write the potential for $F_8 = \dd C_7$
\begin{equation}
\label{eq:F_8_def}
F_8 = \vol(\R^{1,5}) \wedge \dd z \wedge \dd \e^{-4A} \quad \Rightarrow \quad C_7 = \e^{-4A} \vol(\R^{1,5}) \wedge \dd z
\end{equation}
we can reduce \eqref{eq:B_EoM_non_susy_Ima} to
\begin{equation}
\dd (\e^{-2 \phi} * H - m C_7) = -\frac{1}{m} \vol(\R^{1,5}) \wedge \dd \left( \e^{12A - 4 \phi} \dd  ( \partial_z \e^{-4A}) \right)
\end{equation}
which is solved by imposing that $\e^{12A - 4 \phi}$ can be written as a function of $\partial_z \e^{-4A}$:\footnote{If $ \partial_z \e^{-4A}=0$, it is easy to check that the system collapses to the trivial solution $A=$ constant.}
\begin{equation}
\e^{-6A + 2 \phi} = f(\partial_z \e^{-4A}) \, .
\end{equation}
This equation can be used to express $\phi$ in terms of $A$. A direct computation of the off-diagonal components of the Einstein equation now leads to
\begin{equation}
m f ' = \pm 1\,.
\end{equation}
The sign can actually be absorbed by changing the sign of $z$; let us pick the sign $-$ for definiteness. This is solved by $mf = c - \partial_z \e^{-4A}$; in other words,
\begin{equation}
\label{Imamura_Phi_nonsusy}
e^{2\phi} = \frac{e^{6 A}}{m} \left(c-\partial_{z}e^{-4 A} \right) \, .
\end{equation}
Thanks to this equation and \eqref{eq:Eins_0_0}, one gets that all the components of the Einstein and the dilaton equations are automatically satisfied.

We have thus obtained a class of supergravity solutions which generalizes the Imamura class. Let us analyze it a little further. Notice that if $c=0$, we exactly recover the supersymmetric case \eqref{eq:phi_susy}. $H$ is modified as well when $c$ is turned on and in particular it reads
\begin{equation}
\label{eq:Hdef_nonsusy_Ima}
H= \frac{1}{2} \epsilon^{ijk} \dd z \wedge \dd x_i \wedge \dd x_j \partial_k \e^{-6A+2\phi} + \vol(\R^3) ( \partial_z \e^{-10 A+2\phi}- c \, \e^{-6 A+2\phi}) \, .
\end{equation}
Finally, let us consider \eqref{eq:Eins_0_0}; using the definition of $\phi$ we get the following PDE
\begin{equation}
\label{eq:non-SUSY-Ima}
\Delta_3 \e^{-4A}+\frac{1}{2}\partial_{z}^2 \e^{-8A}=- c \left(c-2 \partial_{z} e^{-4 A}\right).
\end{equation}
which is exactly a modification of the Imamura equation \eqref{eq:Ima}. This is the only PDE one has to solve for this non-supersymmetric class.

The parameter $c$ leads to a generalization of the supersymmetric Imamura class of section \ref{sub:imamura}; the supersymmetric case is recovered by setting $c=0$. Since we have found this new class by partially keeping a (bi-)spinorial formalism, it is now possible (even if in principle not necessary) to trace back the supersymmetry-breaking term to a modification of the pure spinor equations. In particular, doing some reverse engineering, we get that the modified pure spinor equations for the non-supersymmetric Imamura solution read:
\begin{subequations}
\label{eq:Imamura_nonSUSY_PureSpinors}
\begin{align}
&\dd_H ( \e^{3A-\phi} \Phi_+) = 0 \, ,\label{eq:Imamura_nonSUSY_PureSpinors1}\\
&\dd_H( \e^{2A-\phi} \Re\Phi_-) = \frac{c}{8} \e^{8A-2\phi} \vol(M_4) \, , \label{eq:Imamura_nonSUSY_PureSpinors2}\\
&\dd_H ( \e^{4A-\phi}\Im\Phi_-) = \frac{\e^{4A}}{8} *_6 \lambda(F) \, . \label{eq:Imamura_nonSUSY_PureSpinors3}
\end{align}
\end{subequations} 
Equation \eqref{eq:Imamura_nonSUSY_PureSpinors3} is unchanged; this was to be expected, since we did not change the expression of the RR fluxes with respect to the BPS class. The only equation which is modified is \eqref{eq:Imamura_nonSUSY_PureSpinors2}, thanks to the introduction of a term proportional to the the volume form of the four-dimensional internal space. This term is controlled by the supersymmetry-breaking parameter $c$. Comparing with (\ref{eq:DW}), we see that from a $d=4$ perspective we can call this an D-term supersymmetry breaking.


 (\ref{eq:Imamura_nonSUSY_PureSpinors}) should be used with care. If we take the same pure spinors $\Phi_\pm$ as in the Imamura class, and we change the definition of $H$ as in (\ref{eq_H_def_non_susy_Ima}), as we have explained, then (\ref{eq:Imamura_nonSUSY_PureSpinors2}) is satisfied, and (\ref{eq:Imamura_nonSUSY_PureSpinors3}) determines all the RR fluxes and hence the modified PDE (\ref{eq:non-SUSY-Ima}).\footnote{In the particular case of this section, it is also true that in fact $H$ is determined by (\ref{eq:Imamura_nonSUSY_PureSpinors2}) uniquely, and that the RR fluxes are in fact unchanged in their functional dependence from $A$ and $\phi$. These latter features are not maintained in our examples, but the general idea remains the same.}


\subsection{Larger IIA system} 
\label{sub:iia_no_S2}

Since the supersymmetric Imamura class is a particular case of the class in section \ref{sub:largerIIA}, one can wonder if it is possible to break supersymmetry also for the latter. 

As we mentioned earlier, the Imamura class is contained in the class of section \ref{sub:largerIIA} up to two T-dualities. Since $\vol(M_4)$ in \eqref{eq:Imamura_nonSUSY_PureSpinors2} does not contain any legs along the six-dimensional external space, the effect of the T-dualities is to transform $\vol(M_4)$ into $\vol(M_6)$. For now, we will simply assume that (\ref{eq:Imamura_nonSUSY_PureSpinors}) is modified to
\begin{subequations}
	\label{eq:IIA_nonSUSY_PureSpinors}
	\begin{align}
	&\dd_H ( \e^{3A-\phi} \Phi_+) = 0 \, ,\\
	&\dd_H( \e^{2A-\phi} \Re\Phi_-) = \frac{h}{8} \e^{8A-4 \phi} \vol(M_6) \, , \label{eq:IIA_nonSUSY_PureSpinors2} \\
	&\dd_H ( \e^{4A-\phi}\Im\Phi_-) = \frac{\e^{4A}}{8} *_6 \lambda(F) \, , \label{eq:IIA_nonSUSY_PureSpinors3} 
	\end{align}
\end{subequations} 
remaining agnostic about the function $h$. The factor $\e^{8A-4 \phi}$ is chosen so that the NSNS three-form reads, once one has solved \eqref{eq:IIA_nonSUSY_PureSpinors}:
\begin{equation}\label{eq:H_larger}
H= - \frac{1}{2} \epsilon^{ijk} \dd z \wedge \dd x_i \wedge \dd x_j \partial_k f + \vol(\R^3) (h+\e^{-\mu}  \partial_z (\e^{-4A-\mu} f)) \, ,
\end{equation}
so that the supersymmetry-breaking function $h$ can be directly seen as a modification of the $\mathrm{vol}(\R^3)$ term in (\ref{eq:susyIIA_H}). 

Since \eqref{eq:IIA_nonSUSY_PureSpinors} doesn't introduce any variation in the one- and two form equations, the metric is unchanged compared to the supersymmetric case \eqref{eq:metIIA_noS2}. The RR fluxes can be easily derived from \eqref{eq:IIA_nonSUSY_PureSpinors3}; the only modification is
\begin{equation}
F_2 = \left( \partial_{y_2}\e^{\mu} \dd y_1 -\partial_{y_1}\e^{\mu} \dd y_2 \right) \wedge \dd z - f^{-1}\left( \partial_z \e^{-4A} + \e^{2 \mu} f^{-1} h \right) \vol(\R^2) \, ,
\end{equation}
while $F_0$ and $F_4$ remain as in \eqref{susyIIA_fluxes}.

Let us now impose the Bianchi identities. Imposing $\dd H = \dd_H F = 0$ strongly constrains the functional form of $h$:
\begin{equation}\label{eq:hcx1}
h = f \e^{\mu} c(x_1) \,, 
\end{equation}
as well as giving the PDEs 
\begin{subequations}\label{eq:larger-sb-PDE1}
	\begin{align}
	&\partial_{x_i} \left( f^{-1} \left(\partial_z \e^{-4A} + c \e^\mu \right) \right) =0 \, , \\
	&\partial_{y_i} \left(\e^{-\mu} \left(\partial_{z} (f \e^{-4A-\mu})+c f \right)\right) =0 \, ,
	\end{align}
\end{subequations}
which modify \eqref{eq:largerIIA_BI1}, and
\begin{subequations}\label{eq:larger-sb-PDE2}
	\begin{align}
	&\triangle_2 \e^{\mu} +\partial_{z} \left( f^{-1} \left(\partial_z \e^{-4A} + c \e^\mu \right) \right) =0 \, , \\
	&\triangle_3 f + \partial_z \left(\e^{-\mu} \left(\partial_{z} (f \e^{-4A-\mu})+c f \right) \right) = 0 \, , \\
	&f \triangle_2 \e^{-4A-\mu} + \triangle_3 \e^{-4A} + f^{-1} \partial_{z}\e^{-4A}  \left(\e^{-\mu} \partial_{z} (f \e^{-4A-\mu}) \right) = -c^2 - \e^{4A} f^{-1} c \partial_{z} \left(\e^{-8A-\mu} f\right) \, \label{eq:lastBI_IIA_gen}
	\end{align}
\end{subequations}
which modify \eqref{eq:largerIIA_BI2}. Imposing these PDEs is enough to solve the equation of motion for $H$; however, the Einstein and dilaton equations impose the extra constraint
\begin{equation}\label{eq:hcx2}
	c'=0\,.
\end{equation}
So $c$ in (\ref{eq:hcx1}) now becomes a constant.

Summing up all these conditions and using the expression (\ref{eq:phi_f}) of $\phi$ in terms of $f$ we get that we can rewrite \eqref{eq:IIA_nonSUSY_PureSpinors2} in term of $c$ as follows:
\begin{equation}
\label{eq:second_modified_}
\dd_H( \e^{2A-\phi} \Re\Phi_-) = -\frac{c}{8} \e^{6A-2 \phi} \vol(M_6) \, 
\end{equation}
where $c$ is now constant. Notice that this is exactly \eqref{eq:Imamura_nonSUSY_PureSpinors2} up to two T-dualities.

Let us pause again to stress what (\ref{eq:second_modified_}) means, along the lines of the comment in the last paragraph of section \ref{sub:im-sb}. We have kept the same pure spinors as in the supersymmetric class of section \ref{sub:largerIIA}, but we have changed $H$ as in (\ref{eq:H_larger}), which has been further fixed by its Bianchi identity as in (\ref{eq:hcx1}), (\ref{eq:hcx2}). Now the usual pure spinor equation (\ref{pure_spinors2}) gets modified to (\ref{eq:second_modified_}); the new (\ref{pure_spinors3}) can now be used to derive modified RR fluxes, which eventually lead to (\ref{eq:larger-sb-PDE1}), (\ref{eq:larger-sb-PDE2}).

Let us also stress that our Ansatz for $H$ was not the most general one, in both the supersymmetric and non-supersymmetric case. For example, one can introduce an additional term governed by the parameter $g$ as following:
\begin{equation}
\label{eq:H_eq_mod}
H= - \frac{1}{2} \epsilon^{ijk} \dd z \wedge \dd x_i \wedge \dd x_j \partial_k f + \vol(\R^3) (h+\e^{-\mu}  \partial_z (\e^{-4A-\mu} f))-g \dd x_1 \wedge \vol (\R^2) 
\end{equation}
and $g$ must be a constant in order to satisfy the Bianchi identity for $H$. Now $F_2$ becomes
\begin{equation}
F_2 = \left( \partial_{y_2}\e^{\mu} \dd y_1 -\partial_{y_1}\e^{\mu} \dd y_2 \right) \wedge \dd z - f^{-1}\left( \partial_z \e^{-4A} + \e^{2 \mu} f^{-1} h \right) \vol(\R^2) + g \dd x_2 \wedge \dd x_3 \, ,
\end{equation}
while the Bianchi identities are all untouched except for \eqref{eq:lastBI_IIA_gen}, which reads:
\begin{equation}
f \triangle_2 \e^{-4A-\mu} + \triangle_3 \e^{-4A} + f^{-1} \partial_{z}\e^{-4A}  \left(\e^{-\mu} \partial_{z} (f \e^{-4A-\mu}) \right) = -c^2- g^2 - \e^{4A} f^{-1} c \,\partial_{z} \left(\e^{-8A-\mu} f\right)  \, .
\end{equation}
Imposing the Bianchi identities (while keeping $c$ constant), one gets that all the EoMs are satisfied. Of course the introduction of the extra parameter $g$ also induces a modification of the pure spinor equations, and in particular the second one becomes:
\begin{equation}
\label{eq:second_pure_spinor_mod}
\dd_H( \e^{2A-\phi} \Re\Phi_-) = -\frac{c-g}{8} \e^{6A-2 \phi} \vol(M_6) \, .
\end{equation}
Notice that by setting $c=g$ we can also choose to restore supersymmetry.


\subsection{General $\R^{1,3} \times S^2$ system} 
\label{sub:iia}

Inspired by the previous cases, let us see if it is possible to extend the supersymmetry breaking procedure to other classes in type IIA supergravity.

The presence of an identity structure has played an important role in the previous discussion, since it allowed to explicitly compute the EoMs and use them to constrain the solution.

Natural candidates are the backgrounds contained in the $\R^{1,3}$ classification of \cite{macpherson-t,apruzzi-geipel-legramandi-macpherson-zagermann}, where the internal space is taken to be a warped product of a two-dimensional sphere with an unconstrained four-dimensional manifold, $M_6 = S^2 \times M_4$. In \cite{l-macpherson} it was proven that the complete list of possible classes of this type in type II theories can be obtained, through a web of duality, starting from two master classes: a conformal Calabi--Yau case in type IIB, which we analyze in the next section, and the IIA solution discussed in \cite[App.~C]{macpherson-t}.

We did not describe this class in section \ref{sec:susy} because it has already appeared in \cite{macpherson-t,apruzzi-geipel-legramandi-macpherson-zagermann}; so let us quickly review it here. Just like for the Imamura case, it is generated by an SU(2)-structure where the vielbein in \eqref{eq:dynamicSU(2)} is given by
\begin{equation}
E_1 = \ii w \, , \qquad E_2 = - \e^C \dd ( \alpha_1+ \ii \alpha_2)+ (\alpha_1 + \ii \alpha_2) v_2 \, , \qquad  E_3 = \e^C \dd \alpha_3 - \alpha_3 v_2 + \ii v_2 \, ,
\end{equation} 
where $w = w_1 + \ii w_2$ and $v = v_1 + \ii v_2$ are a complex vielbein on $M_4$ while $\alpha_i$ are the embedding coordinates of $S^2$ in $\R^3$, which must satisfy $\alpha_1^2+\alpha_2^2 + \alpha_3^2 = 1$. With this choice, the internal metric reads
\begin{equation}
\dd s^2 (M_6) = \e^{2 C} \dd s^2 (S^2) + \dd s^2 (M_4) \, .
\end{equation}

Inserting this expression in \eqref{pure_spinors} one gets that the identity structure on $M_4$ can again be rewritten in terms of local coordinates:
\begin{equation}
v_1= \e^{2A-\phi} \dd z + B_0 \e^{-2A+\phi} \dd r\,,\qquad v_2=- \e^{-2A+\phi}\dd r\,,\qquad  w= \e^{-A}(\dd y_1+i \dd y_2) \, ,
\end{equation}
and the function $C$ is given by
\begin{equation}
r = \e^{2A+C-\phi} \, .
\end{equation}
The names of these coordinates have been chosen so as to make contact with the notation in previous subsections, with $r$ being the radius of $\R^3_x$.

These equations imply the following form of the metric:
\begin{equation}
\label{eq:mink4xS2_metric_IIA}
\begin{split}
\dd s^2=& \e^{2A} \dd s^2(\mathbb{R}^{1,3}) + \e^{-4A+2\phi}\Big(\dd r^2+ r^2 \dd s^2(S^2)\Big)+ \e^{-2A}\Big(\dd y_1^2+ \dd y_2^2\Big) \\
+&  \e^{4A-2\phi}\Big(\dd z+ B_0 \e^{-4A+2\phi}\dd r\Big)^2 .
\end{split}
\end{equation}
The NSNS two-form potential $B$ can be written as
\begin{equation}\label{eq:B0}
B=r^2 \e^{-4A+2\phi}B_0\text{Vol}(S^2) \, ,
\end{equation}
where $B_0$ should satisfy
\begin{subequations}\label{eq: BPS ba10}
	\begin{align}
	&\partial_{r}\left(\e^{2A-2\phi}\right)=\partial_{z}\left(\e^{-2A}B_0\right) \, , \label{eq: BPS ba10 1}\\
	&\partial_{r}\left(r^2\e^{-2A}B_0\right)= \partial_{z}\left(r^2\e^{-6A+2\phi}(1+ B_0^2)\right) \, . \label{eq: BPS ba10 2}
	\end{align}
\end{subequations}
The RR-fluxes are given by
\begin{align}
F_0 =& 0 \, , \nonumber\\[2mm]
F_2=&\big(\partial_{y_2}(\e^{2A-2\phi})\dd y_1-\partial_{y_1}(\e^{2A-2\phi})\dd y_2\big)\wedge \dd z -\partial_{z}(\e^{-4A})\dd y_1\wedge \dd y_2 \nonumber\\
&+\big(\partial_{y_2}(\e^{-2A}B_0)\dd y_1-\partial_{y_1}(\e^{-2A}B_0)\dd y_2\big)\wedge \dd r \, , \\[2mm]
F_4=&B\wedge F_2+r^2\bigg[-\partial_{r}(\e^{-4A})\dd y_1\wedge \dd y_2 -\big(\partial_{y_2}(\e^{-2A}B_0)\dd y_1-\partial_{y_1}(\e^{-2A}B_0)\dd y_2\big)\wedge \dd z \nonumber\\
&+\big(\partial_{y_2}(\e^{-6A+2\phi}(1+B_0^2))\dd y_1-\partial_{y_1}(\e^{-6A+2\phi}(1+B_0^2))\dd y_2\big)\wedge \dd r \bigg]\wedge\text{Vol}(S^2) \, .\nonumber
\end{align}
Finally the Bianchi identities for the fluxes impose the following PDEs:
\begin{align}\label{eq: 4dbianchis}
&\partial^2_{y_1} (\e^{2A-2\phi})+\partial^2_{y_2} (\e^{2A-2\phi})+\partial^2_{z}(\e^{-4A})=0 \, ,\nonumber\\
&\partial^2_{y_1} (\e^{-2A}B_0)+\partial^2_{y_2} (\e^{-2A}B_0)+\partial_{z} \partial_{r}(\e^{-4A})=0 \, ,\\
&\partial^2_{y_1}(r^2 \e^{-6A+2\phi}(1+B_0^2))+\partial^2_{y_2}(r^2 \e^{-6A+2\phi}(1+B_0^2))+ \partial_{r}(r^2\partial_{r}(\e^{-4A}))=0 \, .\nonumber
\end{align}

We now try to break supersymmetry in the supersymmetric class we have just reviewed. Following the strategy in the previous subsections, we again choose to impose all the one- and two-form conditions which determine the identity structure, and therefore we fix the metric to be as \eqref{eq:mink4xS2_metric_IIA}. Moreover, taking inspiration from the modified pure spinor system in the Imamura class \eqref{eq:Imamura_nonSUSY_PureSpinors}, we impose equations \eqref{pure_spinors1} and \eqref{pure_spinors3}, which are enough to fix fluxes and moreover impose the first BPS condition of \eqref{eq: BPS ba10}. Thanks to these constraints, we have that only the six-form part of \eqref{pure_spinors2} is undetermined, just like in \eqref{eq:IIA_nonSUSY_PureSpinors}.

This equation is necessary if one wants to impose supersymmetry, but it is not needed to solve EoMs. 
We claim that we can get a solution, in general non-supersymmetric, just by imposing the Bianchi identities and the pure spinor equations without the six-form part of \eqref{pure_spinors2}, provided a certain condition on $H$ is satisfied.

The proof of this requires heavy computations, which we will not report here. The  condition on $H$ is that it should have at least a leg along $\dd y_1 , \dd y_2$: in other words, recalling (\ref{eq:B0}), 
\begin{equation}\label{eq:B0comp}
	\partial_{y_1,y_2} (\e^{-4A+2\phi}B_0)
\end{equation}
should not be both zero.  Let us sketch an explanation of why this condition is needed. Of course it is not possible to solve in general Bianchi identities and the first BPS condition \eqref{eq: BPS ba10} since, even in the supersymmetric case, this class contains a lot of possible and different subclasses (for example all the AdS$_6$ and AdS$_7$ solutions up to T-dualities). One can try to re-express the EoMs in terms of the Bianchi identities, since both are a system of second order PDEs in terms of the functions $A,\phi,B_0$. For example, it is easy to check that the Bianchi identities imply the EoM for $B$; but not the ones for the dilaton and the metric. However, the Bianchi identities do imply some third-order consistency conditions, which can be obtained by deriving the Bianchi themselves and the first of \eqref{eq: BPS ba10}. 
In fact we found that, in a particular linear combination of these third-order equations, all third order derivatives cancel; this gives an equation of the form 
\begin{equation}\label{eq:aS}
	   (a_1\partial_{y_1} (\e^{-4A+2\phi}B_0)+a_2\partial_{y_2} (\e^{-4A+2\phi}B_0)) S =0\,,
\end{equation}
where $S$ only contains first and second derivatives and $a_1$, $a_2$ are arbitrary functions. It turns out that the Bianchi equations together with $S=0$ now do imply the remaining EoMs. However, if (\ref{eq:B0comp}) are both zero, then the consistency constraint (\ref{eq:aS}) is automatically satisfied without imposing $S=0$.       

To summarize, the system 
\begin{subequations}\label{eq:nonsusy-big}
	\begin{align}
	&\dd_H ( \e^{3A-\phi} \Phi_+) = 0 \, ,\\
	&\dd_H( \e^{2A-\phi} \Re\Phi_-) = h \vol(M_6) \, , \\
	&\dd_H ( \e^{4A-\phi}\Im\Phi_-) = \frac{\e^{4A}}{8} *_6 \lambda(F) \, , 
	\end{align}
\end{subequations}
where this time $h$ is a general function, together with the Bianchi identities, implies all the equations of motion, provided (\ref{eq:B0comp}) are not both zero. 

One might get the impression that this result is in tension with the discussion in section \ref{sub:iia_no_S2}. While there supersymmetry breaking was regulated by the constant $c$ in (\ref{eq:second_modified_}), in (\ref{eq:nonsusy-big}) we have the freedom of a function $h$. But notice that, even if in the discussion of section \ref{sub:iia_no_S2} we impose a rotational symmetry in the $\R^3_{x}$ directions, we do not get a particular case of the discussion of this section. Indeed $H$ in  (\ref{eq:H_larger}) has no legs along $\R^2_{y}$, unlike $\dd$(\ref{eq:B0}). This manifests itself in the fact that in section \ref{sub:iia_no_S2} the equations of motion are not all implied by the Bianchi identities, unless $c$ is a constant.  

It would be interesting to understand if (\ref{eq:nonsusy-big}) is valid also without the presence of the $S^2$, again with the condition that (\ref{eq:B0comp}) is non-zero. However the introduction of the term $g$ in \eqref{eq:H_eq_mod}, which explicitly breaks the $S^2$ isometries, doesn't allow to relax in general the six-form condition but just the specific modification in equation \eqref{eq:second_pure_spinor_mod}. This suggests that to generalize this result without assuming the presence of the $S^2$ some other condition on $H$ is needed.

We do not give a detailed analysis of (\ref{eq:nonsusy-big}) here; imposing the Bianchi identities gives rise to a plethora of possibilities. For the remaining part of this section, we focus on an example. We will explicitly turn on a supersymmetry breaking term and we will show how this breaks the BPS conditions and changes the Bianchi identities. In principle any variation is allowed, but not all of them lead to a nice-looking solution. 

Taking again inspiration from \eqref{eq:Imamura_nonSUSY_PureSpinors},\footnote{In particular, from the fact that the Imamura solution with spherical symmetry can be obtained from this class after two T-dualities \cite{l-macpherson}.} let us postulate
\begin{subequations}
\label{nonsusy_purespinor_IIA}
\begin{align}
&\dd_H ( \e^{3A-\phi} \Phi_+) = 0 \, ,\\
&\dd_H( \e^{2A-\phi} \Re\Phi_-) = \frac{c}{8} \e^{6A-2\phi} \vol(M_6) \, , \\
&\dd_H ( \e^{4A-\phi}\Im\Phi_-) = \frac{\e^{4A}}{8} *_6 \lambda(F) \, .
\end{align}
\end{subequations} 
As we anticipated,  \eqref{eq: BPS ba10 2} is modified by the introduction of a supersymmetry-breaking term:
\begin{equation}
	\frac{1}{x^2_2} \partial_{r} (r^2 \e^{-2A} B_0) = \partial_{z} (\e^{-6A+2 \Phi}(1+B_0^2 )) -c \, .
\end{equation}
Even if the pure spinor equation which defined the fluxes is untouched, the modification of the BPS condition eventually leads to a different expression also for the RR-field
\begin{equation}
\begin{split}
F_0&=0 \, . \\
F_2 &= (\partial_{y_2}\e^{2A-2\phi} \dd y_1 - \partial_{y_1}\e^{2A-2\phi} \dd y_2  ) \wedge \dd z+\big( \partial_{y_2}(\e^{-2A} B_0) \dd y_1  \\
\qquad \qquad &- \partial_{y_1}(\e^{-2A} B_0) \dd y_2\big) \wedge\dd r- \big(\partial_{z}(\e^{-4A})-c \e^{2A-2 \phi} \big)\dd y_1 \wedge \dd y_2 \, ,\\
F_4 &=B\wedge F_2-r^2\Big(  \big( \partial_{y_2}(\e^{-2A} B_0) \dd y_1 - \partial_{y_1}(\e^{-2A} B_0) \dd y_2\big) \wedge\dd z\\
\qquad \qquad &+ \big(\partial_{y_2}(\e^{-6A+2 \phi}(1+B_0^2)) \dd y_1 - \partial_{y_1}(\e^{-6A+2 \phi}(1+B_0^2)) \dd y_2\big) \wedge \dd r \\
\qquad \qquad &-\big(\partial_{r}(\e^{-4A})-c \e^{2A-2 \phi} \big)\dd y_1 \wedge \dd y_2 \Big)\wedge \text{Vol}(S^2) \, ,
\end{split}
\end{equation}
and, as a consequence, of the Bianchi identities:
\begin{subequations}
	\begin{align}
	&\partial^2_{y_1} (\e^{2A-2\phi})+\partial^2_{y_2} (\e^{2A-2\phi})+\partial^2_{z}(\e^{-4A})=c \,   \partial_{z}\e^{2A-2\phi},\\[2mm]
	&\partial^2_{y_1} (\e^{-2A}B_0)+\partial^2_{y_2} (\e^{-2A}B_0)+\partial_{z} \partial_{r}(\e^{-4A})=c\, \partial_{r}\e^{2A-2\phi},\\[2mm]
	&\partial^2_{y_1}(r^2 \e^{-6A+2\phi}(1+B_0^2))+\partial^2_{y_2}(r^2 \e^{-6A+2\phi}(1+B_0^2))+ \partial_{r}(r^2\partial_{r}(\e^{-4A}))=c \, \partial_{r} (r^2 \e^{-2A}B_0).
	\end{align}
\end{subequations}

We will not attempt a detailed analysis of these PDEs, but hopefully they do illustrate that there are many possible ways of breaking supersymmetry even in the master class discussed in \cite[App.~C]{macpherson-t}.


\subsection{Comparison with the conformal Calabi--Yau class} 
\label{sub:gkp}

What we have done in the previous sections has a famous analog in the type IIB conformal Calabi--Yau class \cite{grana-polchinski,giddings-kachru-polchinski,becker2,dasgupta-rajesh-sethi}. In this section we will review this class using the bi-spinorial formalism following \cite{lust-marchesano-martucci-tsimpis} and we will show how it relates to the supersymmetry-breaking technique we have seen so far.

The conformal Calabi--Yau case is obtained by taking an SU(3)-structure, and hence by fixing $k_\perp = 0$ and $k_\parallel= - \ii$ in \eqref{eq:dynamicSU(2)}. Defining  
\begin{equation}
\label{eq:conformal_structures def}
\Omega = \e^{3A - \phi} E_1 \wedge E_2 \wedge E_3 \, , \qquad J = \frac{\ii}{2} \e^{2A - \phi} \left( E_1 \wedge \overline{E}_1 +  E_2 \wedge \overline{E}_2 +  E_3 \wedge \overline{E}_3\right) \, ,
\end{equation}
the pure spinor equations \eqref{pure_spinors} reduce to:
\begin{equation}
\label{eq:GKP_susy}
\begin{split}
&\dd J = \dd \Omega = 0 \, , \qquad \qquad \qquad \quad \, \, \, \, H \wedge \Omega = H \wedge J = 0 \, , \\
& *_6 f_1 = - \frac{1}{2} \e^{-4A} \dd (\e^{\phi} J^2) \, , \qquad *_6 f_3 = \e^{-\phi} H \, , \qquad *_6 f_5 = \e^{-4A} \dd \e^{-4A-\phi} \, .
\end{split}
\end{equation}
The conditions $\dd \Omega = 0$ and $\dd J = 0$, together with the compatibility condition $J \wedge \Omega = 0$, imply that the internal manifold is K\"ahler. However notice that the Calabi--Yau condition is not met due to a conformal factor: $J^3 = \ii \e^{-\phi} \frac{3}{4} \Omega \wedge \overline \Omega$. Yau's theorem still implies the existence of a solution, however.

Let us rewrite the three-form field using the SL$(2,\R)$ covariant formalism and define
\begin{equation}
G = f_3 - \ii \e^{-\phi} H \, .
\end{equation}
Thanks to \eqref{eq:GKP_susy}, we see that $G$ is imaginary self-dual:
\begin{equation}
\label{eq:GKP self}
*_6 G = \ii G \, .
\end{equation}

The action of the Hodge-star operator can be worked out using standard SU(3)-structure identities:
\begin{equation}
*_6 \Omega = - \ii \Omega \, , \qquad *_6 \alpha^0_{(2,1)} = \ii \alpha^0_{(2,1)} \, , \qquad *_6 (\alpha_{(0,1)} \wedge J) = \ii \alpha_{(0,1)} \wedge J \, ,
\end{equation}
where the superscript $^0$ indicates that the form is primitive ($\alpha^0_{(2,1)} \wedge J=0$) while the subscript $_{(m,n)}$ indicates the number of holomorphic and anti-holomorphic components respectively.
Since $\{\overline{\Omega}, \alpha^0_{(2,1)}, \alpha_{(0,1)} \wedge J\}$ span the space of all possible imaginary-self-dual three-forms, $G$ must be a linear combination of these three. However, the supersymmetry constraints $H \wedge J = H \wedge \Omega = 0$ in \eqref{eq:GKP_susy} imply that the components proportional to $\overline{\Omega}$ and $\alpha_{(0,1)} \wedge J$ must be set to zero, which means that $G$ is $(2,1)$ and primitive in order to preserve supersymmetry.

However, in \cite{giddings-kachru-polchinski}, it was discovered that the imaginary-self-duality condition \eqref{eq:GKP self} is enough to drop $G$ out from all the EoM, and therefore we can find non-supersymmetric solutions just by adding to $G$ a $(0,3)$ or a $\alpha_{(0,1)} \wedge J$ component:
\begin{equation}\label{eq:G-sb}
	G= G_\mathrm{BPS} + g \,\bar\Omega + \alpha_{(0,1)} \wedge J\,.
\end{equation}
In terms of pure spinor equations, these two additional components arise from relaxing the six-form part of \eqref{pure_spinors1} and the five-form part of \eqref{pure_spinors2}. While the $g \bar \Omega$ possibility has been widely used, the $\alpha_{(0,1)} \wedge J$ component is usually not considered much because on a compact Calabi--Yau there are no harmonic forms of this type, except for the case of a $T^6$ or $T^2 \times K_3$. This second more exotic possibility is more similar to the ones we considered in IIA, since it arises from modifying the top-form of \eqref{pure_spinors2}.



\section{Examples} 
\label{sec:ex}

In this section we will discuss some particular solutions. Specifically, we are mostly interested in compact solutions which overcome the no-go theorem of \cite{maldacena-nunez}; we will see that this will be possible thanks to the presence of localized O-plane sources. We will focus on the Imamura class of section \ref{sub:imamura} and on its supersymmetry-breaking counterpart \ref{sub:im-sb}. In both cases the problem of finding solutions is reduced to a single nonlinear PDE. To the best of our knowledge, there are no general existence and uniqueness theorems for them. We will hence consider a few possible ansaetze.\footnote{Some of the results in this subsection have been obtained in discussions with G.~B.~De Luca. We also thank P.~Tilli for correspondence about the general theory of the relevant equations.}

\subsection{Separation by sum} 
\label{sub:sep-sum}

We reproduce here the Imamura equation (\ref{eq:Ima}) 
\begin{equation}\label{eq:ima}
	\triangle S + \frac12 \partial_z^2 S^2 = 0 \,.
\end{equation}
where we have now defined $S=\e^{-4A}$.

In this section we consider an Ansatz where
\begin{equation}\label{eq:sep-sum0}
	S = S_1(z)+ S_3(x_1,x_2,x_3)\,.
\end{equation}
(\ref{eq:ima}) immediately imposes that $S_1$ should be linear; moreover, (\ref{eq:phi_susy}) fixes its slope in terms of $m$. We define an integration constant by setting $\e^{-6A+2\phi} = g_s^2$, and we obtain
\begin{equation}\label{eq:sep-sum}
\e^{-4A} = - m g_s^2 z + S_3(x_1,x_2,x_3)\,.
\end{equation}
(\ref{eq:ima}) then reduces to
\begin{equation}
\label{eq:susy_ima_sep1}
\triangle_3 S_3 + m^2 g_s^4 = 0 \, .
\end{equation}
If one is limited by SO(3)-invariance as in \cite{imamura}, then $S_3$ should only depend on $r=(x_i x_i)^{1/2}$, and (\ref{eq:susy_ima_sep1}) becomes an ODE; this results in 
\begin{equation}\label{eq:mD6}
\e^{-4A} = 1- m g_s^2 z - \frac{m^2 g_s^4}{6} r^2 + \frac{Q}{2 r} \,.
\end{equation}
This reproduces the solution in \cite[(26)]{janssen-meessen-ortin}, \cite[(5.7)]{imamura}. Strictly speaking this does not solve (\ref{eq:ima}) but its analogue with a delta function in $r=0$. Since (\ref{eq:ima}) arises in fact from the Bianchi identity $dF_2-F_0 H$, this simply signals the presence of a source. 

In light of this, it is tempting to interpret (\ref{eq:mD6}) as a D6 or O6 in presence of $F_0$, perhaps sourced by a far away D8 transverse to $z$. Indeed in \cite{imamura} different copies of (\ref{eq:mD6}) were pieced together, to obtain a solution with a D8 source. The position of that source was curved; this might just be interpreted as an example of the usual bending of branes. However, the interpretation of (\ref{eq:mD6}) does offer a few puzzles: notice for example that $\e^{-4A}<0$ for large $r$. If we take $Q<0$ to describe an O6, $\e^{-4A}$ also becomes negative at small $r$; while the presence of such a ``hole'' is standard for O6-planes, it is less so that its size depends on $z$, as it does in this case.\footnote{Perhaps this should not be cause for concern. The potential uneasiness comes from the fact that in many situations, especially for numerical solutions, one often recognizes the presence of an O6 by the behavior of the solution at the boundary of the hole, rather than at the ``center'' (in this case $r=0$) where $\e^A<0$ and the solution is unphysical.}

Since in section \ref{sub:imamura} we have derived the Imamura equation (\ref{eq:ima}) without the SO(3) symmetry assumption of \cite{imamura}, we can now obtain more general solutions by solving (\ref{eq:susy_ima_sep1}) without that symmetry. While we are at it, we can try to obtain a compact internal space $M_4$. Let us then periodically identify the $x_i\cong x_i + R$ to describe a torus $T^3$, and let us introduce a source $\sigma = 2\pi Q \delta_R(x_1) \delta_R(x_2) \delta_R(x_3)$, where
\begin{equation}\label{eq:delta-comp}
	\delta_R(x)\equiv \sum_{k\in\mathbb{Z}} \delta(x- kR) = \frac1{R^3}\sum_{k \in \mathbb{Z}}\exp\left[\frac{2\pi\ii}R  k x \right]
\end{equation} 
is the delta function on an $S^1$. Then (\ref{eq:susy_ima_sep1}) is modified to
\begin{equation}\label{eq:DS3-comp}
	\triangle S_3 = - m^2 g_s^4 - \sigma\,.
\end{equation}
Integrating this over the $T^3$ we obtain $2\pi Q = -m^2 g_s^4 R^3$; this tells us the source has negative charge, and should be interpreted as an O6-plane. With this constraint, the combination on the right-hand side of (\ref{eq:DS3-comp}) is exactly such as to subtract the zero mode in the sum (\ref{eq:delta-comp}), and the solution to $S_3$ is\footnote{More rigorous expressions for this solution exist; see for example \cite[Sec.~3.2]{andriot-tsimpis-gw} for a recent discussion.}
\begin{equation}
	S_3 = s_0 + \frac{Q}{2\pi R}\sum_{\vec k \in \mathbb{Z}^3-\{\underline{0}\}} \frac1{k^2}\exp\left[\frac{2\pi\ii}R  \vec k \cdot \vec x \right]\,,
\end{equation}
where $k^2\equiv \vec k \cdot \vec k$. We can now make the solution fully compact by taking $z$ to be periodically identified to describe a circle $S^1$, with $z\cong z + 2 z_0$; in $z\in [z,z_0]$ we take the solution as in (\ref{eq:sep-sum}), while in $z\in [-z_0,0]$ we take (\ref{eq:sep-sum}) with $m\to -m$. At $z=0$ and $z=z_0$, the value of $F_0$ jumps; these loci can be interpreted as O8$_\pm$-planes, a bit as in \cite{cordova-deluca-t-ds4}. $F_0=m$ is fixed in the first copy of the solution to be $m=\frac {4-n_\mathrm{D8}}{2\pi}$. 

The solution represents now the backreaction on a $T^4$ with two O8-planes and an O6-plane. The metric is 
\begin{equation}
\label{eq:sep-sum-met}
\dd s^2_{10} = S^{1/2} \dd s^2 (\R^{1,5})+ S^{1/2} g_s^2 \dd z^2+ S^{-1/2} g_s^2 \dd s^2(T^3) \, ,
\end{equation}
reminiscent of a D6/O6 solution in flat space. Notice however that there is a non-zero $H$ flux;  otherwise having an O6-branes without D6-branes would violate the $F_2$ Bianchi identity. Indeed from (\ref{Imamura_fluxes}) we have
\begin{equation}
	  F_0 = m \, ,\qquad F_2 = \frac{1}{2} \epsilon^{ijk} \dd x_i \wedge \dd x_j \partial_{x_k} S_3 \, ,\qquad H = -m g_s^4\vol(T^3)\,.
\end{equation}
Flux quantization for $H$ is $4\pi^2 N\equiv \int_{T^3} H$, and gives $n_0 N=Q$, which is $-2$ for an O6-plane, where $F_0 = m = \frac{n_0}{2\pi}$, as familiar from other solutions with these ingredients.  The parameters $g_s$ and $s_0$ are still free. Taking $g_s \ll 1$ and $s_0 \gg 1$ makes the solution weakly coupled almost everywhere.  

This solution appears in a slightly different guise in \cite{blaback-vanderwoerd-vanriet-williams,blaback-janssen-vanriet-vercnocke}; there it is interpreted as a domain wall for a seven-dimensional compactification relative to a vacuum solution Mink$_7$ of \cite{blaback-danielsson-junghans-vanriet-wrase-zagermann}.


\subsection{Separation by product} 
\label{sub:sep-prod}

Another possibility is to split 
\begin{equation}
  	S= s(z) S_3(x_1,x_2,x_3) \, .
\end{equation}
(\ref{eq:ima}) now implies
\begin{equation}
\triangle_3 S_3 = k S_3^2 \, , \qquad \partial_z^2 s^2 = -2k s \, .
\end{equation}
The equation on $s$ can be solved analytically by exchanging the role of variable and function: we get $z'(s)=-\frac{\sqrt 3 s}{2 \sqrt{c -k s^3}}$ for some constant $c$. For the equation on $S_3$, there is an existence theorem \cite[Sec.~8.5.2]{evans}: a non-zero solution exists, guaranteeing the existence of a solution on a domain $U\subset \mathbb{R}^3$ with boundary condition $S_3=0$ at the boundary $\partial U$. Since we need $S>0$, we have to take $k<0$. 
If $U$ is taken to be a disk, it is also easy to study the solution numerically. Optimistically, such a solution might be interpreted as the presence of an O6 boundary closing the disk into a three-sphere. We will not investigate this further here.


\subsection{Inverse hodograph transformation} 
\label{sub:hodo}

A common way to linearize a nonlinear PDE is to exchange the role of dependent and independent variables.

We will consider an Ansatz where two of the three coordinates of $\R^2$ are isometries, say $x_2$, $x_3$. Then $S=S(x_1,z)$, and (\ref{eq:ima}) reduces to\footnote{(\ref{eq:ima-1d}) can also be obtained by imposing time-independence on the dispersionless KP equation, which plays a geometrical role in \cite{dunajski-mason-tod}.}
\begin{equation}\label{eq:ima-1d}
\partial_{x_1}^2 S + \frac{1}{2}\partial_z^2 S^2 =0\,.
\end{equation}
We now perform the change of variables
\begin{equation}\label{change}
	x_1 \equiv \partial_U V\,, \qquad z \equiv  \partial_S V\,, \, .
\end{equation}
This turns (\ref{eq:ima-1d}) to
\begin{equation}\label{eq:tricomi}
	\partial_S^2 V + S \partial^2_U V = 0\,.
\end{equation}
which is now linear. Notice that now $S$ has become one of the coordinates. (\ref{eq:tricomi}) is known as Tricomi equation. It changes type from hyperbolic in the region $S<0$ to elliptic in the region $S>0$. It plays a role in modeling the transition from subsonic to transonic regime in fluid dynamics. For us $S=\e^{-4A}$ should be positive, but this transition could play a role if one wanted to analytically continue a solution inside the hole of an O6-plane, although this is of dubious physical significance and will not be attempted here.

In the redefinition (\ref{change}), the old coordinates $x_1$ and $z$ are simply the gradient of $V$ with respect to the new variables $S$ and $U$. Such a transformation is often done in the reverse, i.e.~the new variables are taken to be the gradient of the function in the PDE with respect to the old variables; that is called sometime a ``hodograph'' transformation. A very similar trick was done in \cite{lin-maldacena} to linearize the 2d Toda equation $\triangle_2 S+ \partial_z^2 \e^S=0$ which appears in the classification of ${\cal N}=2$ AdS$_5$ solutions \cite{lin-lunin-maldacena}, and in \cite{ward} for the 1d Toda equation. (\ref{eq:tricomi}) is also a limit of the Chaplygin equation, which can be itself derived as a hodograph transform of Euler's equation for an irrotational fluid. 

Solutions to the Tricomi equation (\ref{eq:tricomi}) are easy to find. For example we can use separation of variables to find
\begin{equation}\label{eq:hodo-airy}
	 V= \sin(U/R) \mathrm{Ai}(R^{-2/3}S) \,.
\end{equation} 
There also exist many other solutions that look more elementary; for example we can take an $\e^{-\ii p U}$ instead of the sine and integrate over $p$ to obtain $V = S(9 U^2 + 4 S^3)^{-5/6}$. The similar-looking solution $V = (9 U^2 + 4 S^3)^{-1/6}$ can be found in \cite[eq. (12)]{youm}. There are also many polynomial solutions. 

In the new coordinates the metric reads
\begin{equation}\label{eq:hodo-met}
	\dd s^2 = \frac1{\sqrt S} \left(\dd s^2 (\R^{1,5})+ \frac{1}{F_0} \partial_U^2 V \left( \frac{\dd s^2(T^2)}{(\partial_U^2 V)^2 + S^{-1}(\partial_S \partial_U V)^2} + (S \dd S^2 +  \dd U^2) \right) \right) 
\end{equation}
and the dilaton is determined by
\begin{equation}
	\e^{2\phi} = \frac{S^{-3/2}\partial_U^2 V}{F_0(S(\partial_U^2 V)^2 +(\partial_S \partial_U V)^2)}\,.
\end{equation}

It would be interesting to be able to make (\ref{eq:hodo-met}) compact. One possibility would be to periodically identify $U\sim U + 2\pi R$; this works for example with the solution (\ref{eq:hodo-airy}) above. However at this point it is not too clear how to make $S$ compact. It is natural to try an approach like that of section \ref{sub:sep-sum}, where we glue two copies of one solution along two O8-planes. However (\ref{eq:ima}) was invariant under $z \to 2z_0-z$, whereas now (\ref{eq:tricomi}) is not invariant under $S\to 2 S_0 -S$, as one would need for this strategy to work. There might of course be other ideas to make (\ref{eq:hodo-met}) compact.


\subsection{Breaking supersymmetry} 
\label{sub:ex-br}

In the previous subsections we have considered several strategies to attack (\ref{eq:ima}). Hopefully that demonstrates that there is a potentially rich array of possibilities already in this case alone, which was the simplest where we broke supersymmetry in section \ref{sec:susy-br}. 

Now we are going to apply the same strategy to the supersymmetry-breaking counterpart of (\ref{eq:ima}), which we reproduce here:
\begin{equation}\label{eq:ima-sb}
	\triangle_3 S + \frac12 \partial_z^2 S^2 + c(c - 2 \partial_z S)=0\,,
\end{equation}
where again we defined $S=\e^{-4A}$. 

So we start with the Ansatz (\ref{eq:sep-sum0}): $S_1$ is restricted to be linear, (\ref{Imamura_Phi_nonsusy}) fixes the slope, and we obtain
\begin{equation}\label{eq:sep-sum-sb}
 	S= (c- m g_s^2) z + S_3(x_1,x_2,x_3)\,,
\end{equation}                             
which replaces (\ref{eq:sep-sum}). Now (\ref{eq:ima-sb}) reduces again to (\ref{eq:susy_ima_sep1}), with $c$ canceling out. From here the discussion is similar to the one in section \ref{sub:sep-sum}. For example in the compact case we can keep for $S_3$ the same solution to (\ref{eq:susy_ima_sep1}). What changes is the expression of $\e^{-4A}=S$, which is modified as in (\ref{eq:sep-sum-sb}), and consequently the expression of the dilaton, which is again fixed by $\e^{-6A+ 2 \phi}= g_s^2$. In particular now the coefficient of $z$ in $S$ can be considered independent from $g_s$. 

An interesting particular case which becomes possible with the introduction of the parameter $c$ is 
\begin{equation}
	c= m g_s^2\,.
\end{equation}
Looking at (\ref{eq:sep-sum-met}) and (\ref{eq:sep-sum-sb}), we see that nothing now depends from the coordinate $z$. Hence a Mink$_7$ emerges, and $z$ is an isometric direction.
We can T-dualize along it and reduce to type IIB. We can moreover T-dualize along the two spatial directions $y_1$, $y_2$ of $\R^{1,5}$. After this chain of dualities we have a Mink$_4$ solution where $H=\frac{c^2}{m} \vol (T^3)$ remains the same; $F_0=m$ gives rise to $F_3 = m \dd z \wedge \dd y_1 \wedge \dd y_2$; and $F_2$ and $F_8$ both contribute to $F_5 = (1+*) \vol(\R^{1,3}) \wedge \dd \e^{-4A}$, where the Hodge star must be taken over the full ten-dimensional space-time. These  fluxes are exactly the ones of the conformal Calabi--Yau class, as one can see from \eqref{eq:GKP_susy}. Supersymmetry is broken by the $J \wedge \alpha_{0,1}$ term we saw in (\ref{eq:G-sb}).

Unfortunately the strategies of sections \ref{sub:sep-prod} and \ref{sub:hodo} don't work for (\ref{eq:ima-sb}), but they were already less successful for the supersymmetric case (\ref{eq:ima}).



\section*{Acknowledgements}

We would like to thank G.B.~De Luca, G.~Ortenzi, P.~Tilli for discussions. We are supported in part by INFN and by MIUR-PRIN contract 2017CC72MK003.

\bibliography{at}
\bibliographystyle{at}

\end{document}